# Infrared spectra of $Rg_{1,2}$ - $C_6H_6$ complexes, Rg = He, Ne, Ar

K. Esteki,[a] A.J. Barclay,[a] A.R.W. McKellar,[b] N. Moazzen-Ahmadi [a,*]

[a] Department of Physics and Astronomy, University of Calgary, 2500 University Drive North West, Calgary, Alberta T2N 1N4, Canada

[b] National Research Council of Canada, Ottawa, Ontario K1A 0R6, Canada

Address for correspondence:     Dr. N. Moazzen-Ahmadi
                                Department of Physics and Astronomy,
                                University of Calgary,
                                2500 University Drive North West,
                                Calgary, Alberta T2N 1N4,
                                Canada

* Corresponding author. Fax: +1-403-289-3331; e-mail: nmoazzen@ucalgary.ca (N. Moazzen-Ahmadi)

Abstract

Infrared spectra of $Rg_{1,2}$ - $C_6H_6$ complexes (Rg = He, Ne, Ar) are observed in the region of the $\nu_{12}$ fundamental of $C_6H_6$ using a pulsed supersonic jet expansion and a tunable optical parametric oscillator laser source. The mixed trimer He - Ne - $C_6H_6$ is also detected. Four bands are analyzed for each complex, namely $\nu_{12}$ itself ($\approx 3048$ cm$^{-1}$) and three linked combination bands ($\approx 3079$, $3100$, and $3102$ cm$^{-1}$). The results are consistent with previous ultraviolet and microwave results, with $Ne_2$ - $C_6H_6$ and He - Ne - $C_6H_6$ being analyzed spectroscopically here for the first time.

**Keywords**:    van der Waals, rare gas, benzene, infrared spectrum, supersonic jet, Fermi interaction

## 1. Introduction

There is an extensive history of spectroscopic studies of weakly-bound rare gas (Rg) - benzene clusters [1], ranging from microwave to ultraviolet wavelength regions, and including techniques such as Fourier transform microwave, laser induced fluorescence, coherent ion dip, resonant enhanced two photon ionization, and nonlinear Raman spectroscopies [2-16]. Rg - benzene dimers have simple structures with the rare gas atom localized on the $C_6$ symmetry axis at a distance of about $r_0 = 3.4 - 3.8$ Å (depending on the Rg) from the plane of the benzene molecule. For the $Rg_2$ - benzene trimers, the second rare gas atom takes up an equivalent position on the other side of the benzene. Thus the trimers have $D_{6h}$ point group symmetry similar to benzene monomer (assuming the same Rg atoms), while the dimers have $C_{6v}$ symmetry.

However, there is relatively little work on Rg - benzene spectra in the infrared region, with the only previous report being our study of He -, Ne -, and Ar - $C_6D_6$ [16] in the region of the $\nu_{12}$ C-D stretching band of $C_6D_6$ near 2289 cm$^{-1}$. This involved two bands, the $\nu_{12}$ fundamental itself, and the $\nu_2 + \nu_{13}$ combination band which is strongly coupled to it by an anharmonic (Fermi-type) resonance [17,18]. In the present paper, we extend these $C_6D_6$ results to the analogous C-H stretching band of the principal isotopologue, $C_6H_6$, and also detect the trimers (e.g. $He_2$ - $C_6H_6$) in addition to the dimers. It turns out that the $C_6H_6$ $\nu_{12}$ region is more complicated than that of $C_6D_6$, so that four bands are actually observed, the fundamental plus three linked combination vibrations.

## 2. Results

Spectra were obtained with our previously described pulsed supersonic slit jet apparatus at the University of Calgary, probed with a Lockheed Martin Aculight Argos optical parametric oscillator source [16,18-20]. The expansion gas mixtures contained approximately 0.02% $C_6H_6$ in He carrier gas for He-$C_6D_6$, with the addition of a 0.5% Ne or Ar for Ne-$C_6H_6$ or Ar-$C_6H_6$. Backing pressures were about 9 atmospheres, and etalon and $N_2O$ reference gas spectra were simultaneously recorded for wavenumber calibration. The PGopher computer program [21] was used for spectral simulation and fitting, including the appropriate nuclear spin weights. For $C_6H_6$ monomer and $He_2$-, $Ne_2$-, and $Ar_2$- $C_6H_6$, which have $D_{6h}$ symmetry, these spin weights are: 7 or 3 for $K'' = 0$, $J'' =$ even or odd; and 11 : 9 : 14 : 9 : 11 : 10 for $K'' = 1 + 6n : 2 + 6n : 3 + 6n : 4 + 6n : 5 + 6n : 6 + 6n$, where $n = 0, 1, 2, \ldots$. For $Rg_1$ - $C_6H_6$ and $He - Ne$ - $C_6H_6$, which have $C_{6v}$ symmetry, the spin weight are: 10 for $K'' = 0$; and 22 : 18 : 28 : 18 : 22 : 20 for $K'' = 1 + 6n : 2 + 6n : 3 + 6n : 4 + 6n : 5 + 6n : 6 + 6n$.

As usual for symmetric top molecules, the rotational constant for motion around the symmetry axis is not easily determined by spectroscopy and has little direct effect on other parameters. In the analyses below, we assume that this parameter ($A$ or $C$) for the ground vibrational state of Rg-$C_6H_6$ complexes is equal to that of $C_6H_6$ itself, 0.0948753 cm$^{-1}$ [22].

## 2.1. $C_6H_6$

Our analyses are based on a large and detailed study of the $C_6H_6$ monomer spectrum made by Pliva and Pine [22] involving the $\nu_{12}$ fundamental and 18 perturbing states, which included over 8000 transitions and 112 adjustable parameters. Of these 18 perturbers, 3 are rather strongly coupled to the fundamental and to each other by Fermi-type interactions, with the result that they have band strengths comparable to that of the fundamental itself and are easily observed. This situation for $C_6H_6$ monomer, as determined in [22], is summarized in Table 1, where the four states are labelled A, B, C, and D, following Pliva and Pine. Note the very extensive and somewhat complicated mixing of these four vibrations, which results, for example, in a re-ordering of the energies of the original unperturbed basis states A, B, C, D to a perturbed (observed!) order of A, C, D, B. Note also that perturbed states D and B are separated by only about 1.5 cm$^{-1}$, which means in practice that the two bands overlap each other.

Table 1. The linked $E_{1u}$ vibrations of $C_6H_6$ near 3100 cm$^{-1}$, as analyzed by Pliva and Pine [22]

| Label | Vibrational mode | Deperturbed origin [a] (cm$^{-1}$) | Perturbed (observed) origin [b] (cm$^{-1}$) | Fractional parentage [c] | | | |
|:---:|:---:|:---:|:---:|:---:|:---:|:---:|:---:|
| | | | | A | B | C | D |
| A | $\nu_{12}$ | 3064.3674 | 3047.9080 | **0.65** | 0.32 | 0.03 | 0.00 |
| B | $\nu_{13} + \nu_{16}$ | 3083.7847 | 3101.8849 | 0.15 | **0.44** | 0.07 | 0.34 |
| C | $\nu_2 + \nu_{13} + \nu_{18}$ | 3080.0395 | 3078.8476 | 0.11 | 0.03 | **0.86** | 0.00 |
| D | $\nu_3 + \nu_{10} + \nu_{18}$ | 3100.8602 | 3100.4113 | 0.10 | 0.21 | 0.03 | **0.66** |

[a] The anharmonic resonance (Fermi-type) parameters linking these vibrations are: $W_{AB}$ = 23.0206, $W_{BC}$ = 9.0480, $W_{AC}$ = -1.1110, $W_{AD}$ = -1.0641, $W_{BD}$ = 1.5136, $W_{CD}$ = 0.0 cm$^{-1}$.

[b] Some of our perturbed origins do not agree exactly with the values quoted in the abstract of [22], notably that of C which was given there as 3078.614. But the "real" origin of C given here is indeed correct, as shown by fitting C in isolation which gives an origin of 3078.846.

[c] Fractional parentages are the squares of the eigenvectors of the diagonalized energy matrix.

We observed these same four bands for the Rg - $C_6H_6$ complexes, as described below. For their analysis, we followed the scheme of Pliva and Pine, and kept most of the anharmonic interaction parameters, $W$, fixed at their monomer values from [22].

## 2.2. He - $C_6D_6$ and He$_2$ - $C_6H_6$

In 1979, Beck et al. [2] observed the $S_1 \leftarrow S_0$ $6^1_0$ vibronic band ($\approx$38610 cm$^{-1}$) for both He$_1$ - $C_6H_6$ and He$_2$ - $C_6H_6$ using laser induced fluorescence in a pulsed supersonic jet expansion with an effective rotational temperature of about 0.3 K. More recently, this band was re-examined by Hayashi and Ohshima [3] with improved spectral resolution ($<0.01$ cm$^{-1}$) using resonant enhanced two photon ionization spectroscopy. Their analysis yielded ground state $B$-values of 0.1220 and 0.0876 cm$^{-1}$ for the dimer and trimer, respectively, which resulted in intermolecular distances of $r_0 = 3.602$ or 3.596 Å (these are for the distance between the He atom(s) and the benzene plane). A detailed *ab initio* potential surface for He-$C_6H_6$ has been reported by Lee et al. [10], together with nonlinear Raman spectra of some intermolecular vibrations

We observed He$_{1,2}$ - $C_6H_6$ spectra accompanying all four benzene bands in the 3040 - 3110 cm$^{-1}$ range, as illustrated in Fig. 1 (recall that the B and D bands near 3100 cm$^{-1}$ overlap). The jet expansion conditions resulted in He$_1$ and He$_2$ spectra whose strongest features were not much weaker than those of the $C_6H_6$ monomer itself. This is shown by the simulated spectra in Fig. 1, which are based on our analyses described below. The strongest features are $Q$-branches which contain many transitions which are unresolved due to the low effective rotational temperature ($\approx 2.5$ K) and relatively small changes in $B$-values. Hence the number of assigned lines is considerably less than the number of assigned transitions. For example, the number of assigned lines for He$_1$ - $C_6H_6$ was 191 while the number of transitions was 340. The spacing of these $Q$-branches varies greatly, so that the 'look' of each band also varies. Since this spacing is approximately given by $2[B' - C'(1 - \zeta)]$, the $Q$-branches collapse into a single feature when $\zeta = (1 - B'/C')$, as is (almost) the case for He$_1 - C_6H_6$ at 3101.92 cm$^{-1}$ in the B band (Fig. 1).

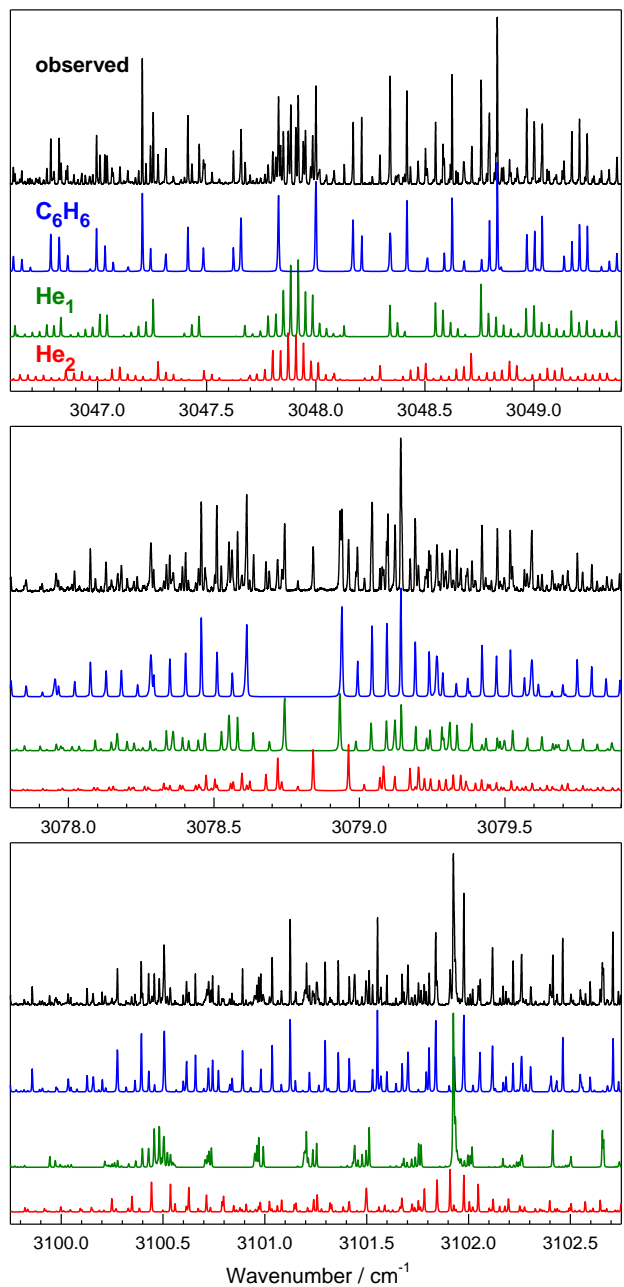

Fig.1. Observed (upper traces) and simulated (lower traces) spectra of He$_1$-C$_6$H$_6$ and He$_2$-C$_6$H$_6$. The simulated spectra are based on the parameters in Table 2 with a temperature of 2.5 K, a Gaussian line width of 0.0034 cm$^{-1}$ and a Lorentzian width of 0.001 cm$^{-1}$.

Table 2. Molecular parameters for $C_6H_6$ and $Rg_{1,2}$-$C_6H_6$ complexes. [a]

| State | Parameter | $C_6H_6$ | $He-C_6H_6$ | $He_2-C_6H_6$ | $Ne-C_6H_6$ | $Ne_2-C_6H_6$ | $Ar-C_6H_6$ [c] | $Ar_2-C_6H_6$ |
|---|---|---|---|---|---|---|---|---|
| G | $B''$ | 0.1897618 | 0.122026(17) | 0.087648(23) | 0.060383616 | 0.029831(14) | 0.0394025766 | 0.015144(11) |
| | $D_J''$ | 3.934e-8 | 3.82(24)e-6 | 2.54(31)e-6 | 0.64658e-6 | | 1.08665e-7 | |
| | $D_{JK}''$ | -6.90e-8 | | | 2.9487e-6 | | 5.937808-7 | |
| A | $v_0$ | 3064.3688(4) | 3064.412(2) | 3064.429(8) | 3064.340(3) | 3064.269(7) | 3064.2674(10) | 3064.153(3) |
| | $B'$ | 0.1896203 | 0.121928(41) | 0.087533(31) | 0.0603744(9) | 0.029830(16) | 0.0394161(42) | 0.015143(12) |
| | $\zeta'$ | -0.07891 | -0.09321(79) | -0.0925(16) | -0.09423(69) | -0.0869(18) | -0.08141(39) | -0.08257(63) |
| | $D_J'$ | 3.45e-8 | 1.81(61)e-6 | | c | | c | |
| B | $v_0$ | 3083.7804(5) | 3083.736(1) | 3083.722(5) | 3083.679(2) | 3083.618(5) | 3083.3404(7) | 3082.938(2) |
| | $B'$ | 0.1893552 | 0.122150(50) | 0.087777(77) | 0.060310(15) | 0.029828(20) | 0.0393203(69) | 0.015143(13) |
| | $\zeta'$ | -0.24145 | -0.2459(14) | -0.2475(27) | -0.2506(14) | -0.2432(34) | -0.25628(74) | -0.2427(12) |
| | $D_J'$ | 4.50e-8 | 9.84(76)e-6 | 7.5(10)e-6 | c | | c | |
| C | $v_0$ | 3080.0391(2) | 3080.100(1) | 3080.168(1) | 3080.048(1) | 3080.072(2) | 3079.8399(2) | 3079.6331(5) |
| | $B'$ | 0.1895099 | 0.121986(18) | 0.087686(28) | 0.0604324(46) | 0.029831(14) | 0.0394622(17) | 0.015145(11) |
| | $\zeta'$ | 0.86042 | 0.84785(29) | 0.84849(41) | 0.84716(23) | 0.84980(37) | 0.85108(10) | 0.84549(27) |
| | $D_J'$ | 3.58e-8 | 4.62(24)e-6 | 2.43(38)e-6 | c | | c | |
| D | $v_0$ | 3100.8623(4) | 3100.881(1) | 3100.901(2) | 3100.817(1) | 3100.781(1) | 3100.6611(4) | 3100.4668(8) |
| | $B'$ | 0.189646 | 0.121758(31) | 0.087524(34) | 0.0603721(74) | 0.029828(15) | 0.0394118(42) | 0.015146(12) |
| | $\zeta'$ | -0.585415 | -0.59939(79) | -0.5979(15) | -0.59990(64) | -0.5996(16) | -0.58771(48) | -0.58847(81) |
| | $D_J'$ | 3.93e-8 | 0.61(40)e-6 | | c | | c | |
| | $W_{AB}$ | 23.0206 | 23.05502(96) | 23.0768(37) | 23.0551(13) | 23.0736(32) | 23.12402(47) | 23.2138(12) |
| | # | | 191 | 120 | 135 | 41 | 138 | 27 |
| | rms | | 0.00083 | 0.00084 | 0.00069 | 0.00083 | 0.00046 | 0.00044 |

[a] In all cases, $C$ (or $A$) was fixed at the following $C_6H_6$ values from Pliva and Pine [22]: 0.0948573 (ground state); 0.0947423 (state A); 0.0946436 (state B); 0.0946220 (state C); 0.0947774 (state D). Similarly, all anharmonic resonance parameters except $W_{AB}$ were fixed: $W_{BC}$ = 9.0480; $W_{AC}$ = -1.1110; $W_{AD}$ = -1.0641; $W_{BD}$ = 1.5136; $W_{CD}$ = 0.0; $\lambda_{AB}$ = -0.5540e-4; $\lambda_{BC}$ = 1.6339e-4 (all values in cm$^{-1}$). $C_6H_6$ monomer parameters are from Table III of Pliva and Pine [22], except that the band origins were fitted to present data, with all other $C_6H_6$ parameters fixed at their Pliva and Pine values (see text). For Ne – and Ar – $C_6H_6$, ground state parameters are from Brupbacher et al. [13], and excited state $D_J'$ and $D_{JK}'$ parameters are fixed at ground state values.

Our analysis was made using the symmetric rotor Hamiltonian in the PGopher program [21]. For He$_1$ - C$_6$H$_6$, the ground state parameters $B''$ and $D_J''$ were varied, as well as the excited state parameters $v_0$, $B'$, $D_J'$, and $\zeta$ for each of the four vibrations A to D. In addition, the largest of the anharmonic interaction parameters, $W_{AB}$, was varied. Meanwhile, many parameters were fixed at C$_6$H$_6$ values [22], including the ground and excited state $C$-values and the remaining interaction parameters, $W$ and $\lambda$. For He$_2$ - C$_6$H$_6$, the situation was the same except that some $D_J$ parameters were not well determined and were fixed to zero. The results of these fits are shown in Table 2.

## 2.3. Ne$_1$ - C$_6$D$_6$, Ne$_2$ - C$_6$H$_6$, and He - Ne - C$_6$H$_6$

Ne$_1$ - C$_6$H$_6$ has been studied previously by electronic [4] and microwave [12,13] spectroscopy, but as far as we know there are no previous reports of Ne$_2$ - C$_6$H$_6$. The present Ne spectra are shown in Fig. 2. In analyzing the Ne$_1$ spectrum, we had the advantage of very precise ground state parameters ($B''$, $D_J''$, and $D_{JK}''$) from the microwave studies [13]. A total of 135 lines were assigned in terms of 380 transitions, and fitted by varying the excited state $v_0$, $B'$, and $\zeta$ parameters for each of the four vibrations, plus the $W_{AB}$ interaction parameter (see Table 2). Other parameters were fixed as for He$_1$ - C$_6$H$_6$, and in addition the excited state $D_J'$, and $D_{JK}'$ parameters were fixed to their ground state microwave values.

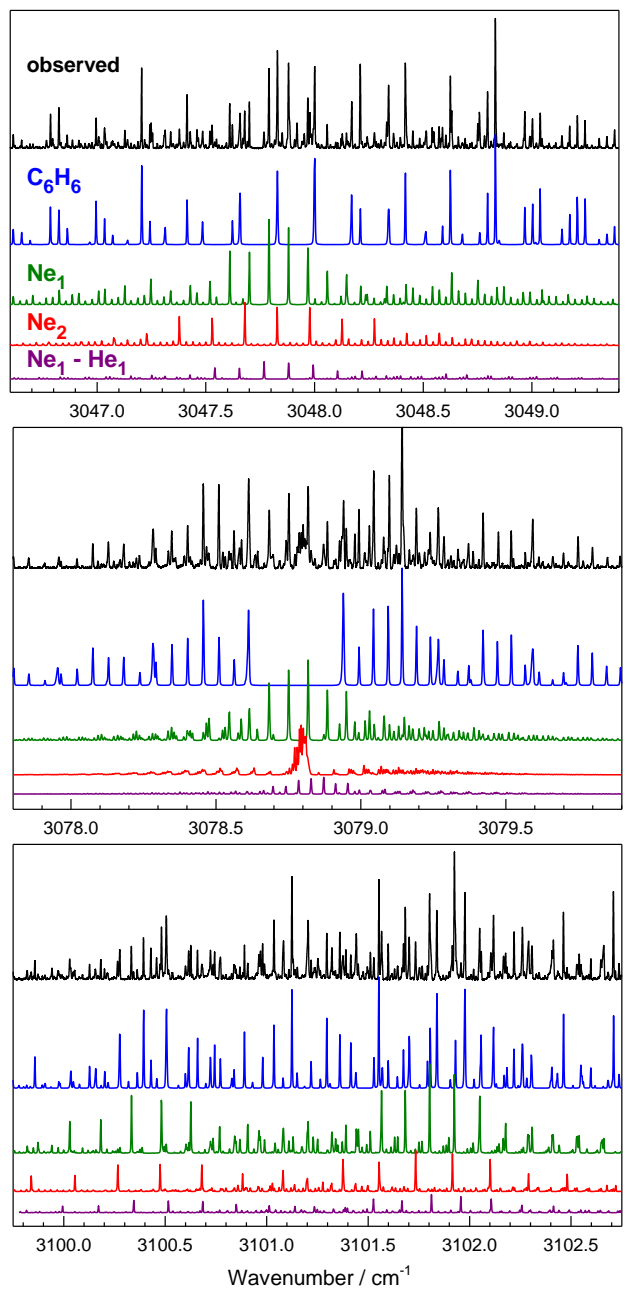

Fig.2. Observed and simulated spectra of Ne$_1$-C$_6$H$_6$, Ne$_2$-C$_6$H$_6$, and He-Ne-C$_6$H$_6$.

For $Ne_2$ - $C_6H_6$, the number of assigned lines was much smaller (41), as can perhaps be understood by looking at Fig. 2. Moreover, there were no previous data for the ground state (from symmetry, $Ne_2$ - $C_6H_6$ has no permanent dipole moment and hence no microwave spectrum). The analysis varied the ground state $B''$, the excited state $v_0$, $B'$, and $\zeta$ for the four vibrations, and $W_{AB}$, with results as shown in Table 2.

As already mentioned, helium was used as the carrier gas in the mixture used to obtain the Ne spectra, relying on the much larger binding energy of Ne to yield mostly $Ne_{1,2}$ - $C_6H_6$ in the resulting supersonic expansion. But in fact $He_1$ - $C_6H_6$ was still present in our spectra, with an intensity roughly 10% that of $Ne_1$ - $C_6H_6$. Furthermore, the mixed trimer He - Ne - $C_6H_6$ was also present and we were able to assign a number of lines to it. The resulting simulated spectra are shown at the bottom of the panels in Fig. 2. Most of the assigned lines were $Q$-branch transitions, which provide almost no information about $B$-values. By assigning some very weak observed $R$-branch transitions, and constraining the ground and excited $B$-values to be equal, we were able to obtain a meaningful, but tentative, value of 0.02983 cm$^{-1}$ for $B$. The He - Ne - $C_6H_6$ analysis is summarized in Table 3.

Table 3. Molecular parameters for He – Ne - $C_6H_6$. [a]

| State | Parameter | |
|-------|-----------|---|
| G | $B''$ | 0.048221(19) |
| A | $v_0$ | 3064.3524(3) |
| | $B'$ | a |
| | $\zeta''$ | -0.0802(12) |
| B | $v_0$ | 3083.6688(6) |
| | $B'$ | a |
| | $\zeta''$ | -0.2581(23) |
| C | $v_0$ | 3080.1194(1) |
| | $B'$ | a |
| | $\zeta''$ | 0.8467(3) |
| D | $v_0$ | 3100.8411(6) |
| | $B'$ | a |
| | $\zeta''$ | -0.5900(16) |
| | $W_{AB}$ | 23.075 [b] |
| | # | 20 |
| | rms | 0.00034 |

[a] Other parameters were fixed as in Table 2, footnote a. Excited state $B$-values were constrained to equal the ground state value, $B''$, which is tentative (see text). The $1\sigma$ statistical uncertainties given here almost certainly underestimate the true uncertainties.

[b] $W_{AB}$ was fixed at this value, which is the average of those of $He_2$ - and $Ne_2$ - $C_6H_6$.

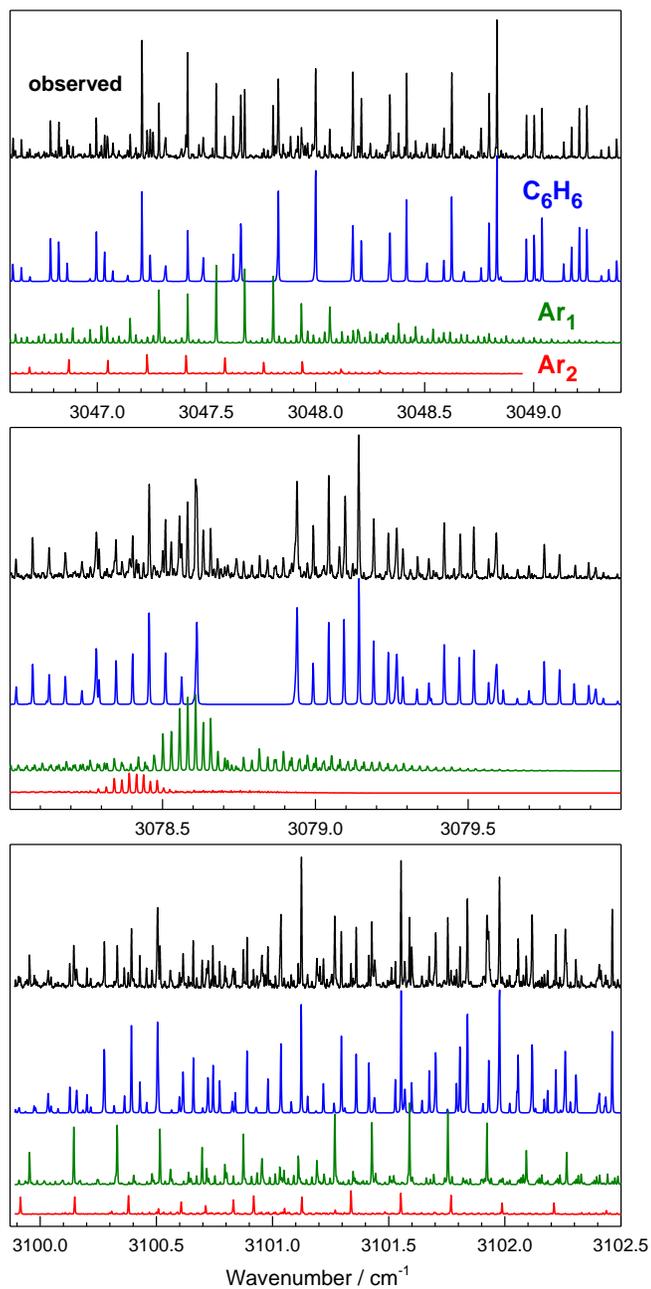

Fig.3.  Observed and simulated spectra of $Ar_1$-$C_6H_6$ and $Ar_2$-$C_6H_6$.

### 2.4. Ar$_1$ - C$_6$H$_6$ and Ar$_2$-C$_6$H$_6$

Similar to the Ne case, Ar-benzene dimers have been extensively studied: a high level *ab initio* calculation and references to previous work are given by Fernández *et al.* [23], and accurate ground state constants ($B$, $D_{JK}$, $D_J$) are available for Ar$_1$ - C$_6$H$_6$ [13]. Our spectra are shown in Fig. 3. For the dimer, Ar$_1$ - C$_6$H$_6$, a total of 138 lines involving 461 transitions were fitted to give the parameters listed in Table 2, varying the same 13 upper state parameters as for Ne$_1$ - C$_6$H$_6$.

For the trimer, Ar$_2$ – C$_6$H$_6$, electronic spectra have been observed previously [5], and a ground state $B''$-value was thus available. But we felt that we could determine $B''$ with comparable accuracy and therefore allowed it to vary in the analysis along with the same 13 upper state parameters as for Ne$_2$ - C$_6$H$_6$, with results as shown in Table 2. There were 27 assigned lines involving 211 transitions. There was some evidence for the presence of the mixed trimer He - Ar - C$_6$H$_6$ in our spectra, but the tentatively assigned transitions were insufficient for a meaningful analysis.

### 3. Discussion

### 3.1. $\zeta$-values, $B$-values, and intermolecular distances

Tables 2 and 3 show that the addition of one or two rare gas atoms to C$_6$H$_6$ has little effect on the Coriolis zeta ($\zeta$) parameters of the individual vibrational modes, just as observed for Rg – C$_6$D$_6$ dimers [16]. This is not surprising, since it simply indicates that the weakly-bound Rg atom does not have much effect on the interactions among the intramolecular benzene vibrations. As noted above (Sec. 2.2), the quantity $2[B - C(1 - \zeta)]$ has a significant effect on the overall appearance of the bands. The parameters $C$ and $\zeta$ remain mostly unchanged with the addition of Rg atoms, but of course the $B$-value, and hence the band appearance, does change significantly.

Our ground state $B$-values for $He_1$- and $He_2$ - $C_6H_6$ are 0.12203(5) and 0.08765(7) $cm^{-1}$, respectively, which agree very well with the slightly less precise values of 0.1220(2) and 0.0876(2) $cm^{-1}$ reported by Hayashi and Ohshima [3] (these are $3\sigma$ uncertainties). In the case of $Ar_2$ - $C_6H_6$, our ground state $B$-value of 0.01514(3) $cm^{-1}$ is slightly lower than value of 0.01517(2) $cm^{-1}$ reported by Weber and Neusser [5]. For $Ne_2$ - $C_6H_6$ and He - Ne - $C_6H_6$ there are no previous $B$-value determinations with which to compare. Finally, for $Ne_1$- and $Ar_1$ - $C_6H_6$ there are very accurate previous microwave $B$-values, as already noted [12,13].

Table 4. Effective intermolecular distances (in Å).

|                    | n = 1 | n = 2 |
|--------------------|-------|-------|
| $He_n$ - $C_6H_6$  | 3.599 | 3.596 |
| $Ne_n$ - $C_6H_6$  | 3.458 | 3.451 |
| $Ar_n$ - $C_6H_6$  | 3.581 | 3.580 |

Effective intermolecular distances (Rg atom to benzene plane) resulting from these $B$-values are summarized in Table 4. Here we use the $C_6H_6$ $B$-value as given in Table 2 [22], and also assume that the rare gas atoms are rigidly located on the benzene symmetry axis. The latter point is discussed in [12], where it is shown that allowing for the off-axis motion of Ne significantly affects ($\approx 0.05$ Å) the intermolecular distance obtained for Ne – $C_6H_6$, but in most cases we do not have isotopic data to evaluate this effect. From Table 4 we see that the distances vary from 3.599 Å for $He_1$, down to 3.458 Å for $Ne_1$, and back up to 3.581 Å for $Ar_1$, due to the competing effects of atomic 'size' (larger for Ar) and zero-point motion (larger for He). The distances become smaller when a second rare gas atom is added, as expected due to the (relatively weak) attractive effects between the two atoms. For Ar, the effect appears to be very small (0.001 Å), but this could be due to uncertainty in our $Ar_2 – C_6H_6$ $B$-value (if the $B$-value from [5] is used, the shrinkage becomes 0.004 Å). We cannot separately evaluate the two intermolecular distances for He - Ne - $C_6H_6$, but the observed $B$-value is consistent with a slight shrinkage compared to He - $C_6H_6$ and Ne - $C_6H_6$.

Table 5. Vibrational shifts (in cm$^{-1}$)

| | Deperturbed | | | | Perturbed (real) | | | |
|---|---|---|---|---|---|---|---|---|
| | A | B | C | D | A | B | C | D |
| He$_1$-C$_6$H$_6$ | 0.043 | -0.044 | 0.061 | 0.019 | -0.015 | 0.060 | 0.019 | 0.015 |
| 2 x He$_1$-C$_6$H$_6$ | 0.087 | -0.088 | 0.121 | 0.038 | -0.030 | 0.119 | 0.039 | 0.030 |
| He$_2$-C$_6$H$_6$ | 0.060 | -0.059 | 0.129 | 0.038 | -0.026 | 0.123 | 0.039 | 0.033 |
| Ne$_1$-C$_6$H$_6$ | -0.029 | -0.101 | 0.009 | -0.045 | -0.082 | 0.005 | -0.044 | -0.047 |
| 2 x Ne$_1$-C$_6$H$_6$ | -0.058 | -0.203 | 0.018 | -0.091 | -0.164 | 0.011 | -0.088 | -0.093 |
| Ne$_2$-C$_6$H$_6$ | -0.100 | -0.163 | 0.033 | -0.081 | -0.165 | 0.019 | -0.081 | -0.085 |
| Ar$_1$-C$_6$H$_6$ | -0.101 | -0.440 | -0.199 | -0.201 | -0.307 | -0.184 | -0.212 | -0.238 |
| 2 x Ar$_1$-C$_6$H$_6$ | -0.203 | -0.880 | -0.398 | -0.402 | -0.614 | -0.368 | -0.425 | -0.477 |
| Ar$_2$-C$_6$H$_6$ | -0.216 | -0.843 | -0.406 | -0.396 | -0.600 | -0.376 | -0.418 | -0.466 |
| He$_1$ + Ne$_1$ | 0.014 | -0.146 | 0.069 | -0.026 | -0.097 | 0.065 | -0.024 | -0.032 |
| He$_1$-Ne$_1$-C$_6$H$_6$ | -0.016 | -0.112 | 0.080 | -0.021 | -0.137 | 0.076 | -0.007 | -0.001 |

**Vibrational shifts**

The most novel results of the present study are perhaps the vibrational shifts, that is, the differences between the band origins of the various Rg – benzene clusters and those of benzene itself. However, ready interpretation of these shifts is made a bit difficult by the complicated nature of the four coupled benzene vibrations as described in Table 1. The observed shifts are summarized in Table 5, which give values for both the deperturbed and perturbed (observed) origins of bands A, B, C, and D. For each Rg atom, we list the shift for Rg$_1$ - and Rg$_2$ - C$_6$H$_6$ as well as twice the value of the Rg$_1$ shift. In most cases, the Rg$_2$ shift is quite similar to twice the Rg$_1$ shift, as expected, and this is especially true for the 'observed' origins, which are probably more meaningful. For the mixed trimer, He - Ne - $C_6H_6$, we similarly compare shifts with the sum of those of $He_1$ and $Ne_1$ and also find rough agreement.

The Ar shifts tend to be more negative (red shift) and larger in magnitude than those of He, with Ne falling in between. This trend is similar to many other weakly-bound complexes, for example Rg - CO [24]. Looking at the individual bands (observed origins), we find that band A is invariably the most red-shifted, and band B the least red-shifted, with C and D being intermediate and similar to each other. Table 5 contains rather subtle information about interactions between the Rg atoms and the various benzene vibrations, but any further interpretation is difficult at this point.

## 4.   Conclusions

In conclusion, we report here the first spectroscopic investigation of rare gas – $C_6H_6$ complexes in the infrared region, complementing our earlier study of Rg – $C_6D_6$ [16]. Both $Rg_1$ and $Rg_2$ complexes were observed for Rg = He, Ne, and Ar, and the mixed trimer He - Ne - $C_6H_6$ was also detected. In all cases, the Rg atoms are located on the $C_6$ symmetry axis at distances of about 3.5 Å from the plane of the benzene molecule. More precisely, the observed $B$-values give intermolecular distances of about 3.60, 3.45, and 3.58 Å for He, Ne, and Ar, respectively. The somewhat complicated pattern of four bands linked with the $v_{12}$ vibration of $C_6H_6$ is preserved virtually intact in the Rg complexes, with vibrational shifts of <1 cm$^{-1}$ (<0.03%). As well, the Coriolis zeta parameters of the four modes remain little changed in the complexes.

**Acknowledgments**

The authors gratefully acknowledge the financial support of the Natural Sciences and

Engineering Research Council of Canada.